\documentclass{emulateapj} 
\usepackage{graphicx}
\usepackage{amsmath}
\begin{document}
\title{Quark matter in massive compact stars}
\author{Simon Weissenborn$^1$, Irina Sagert$^2$, Giuseppe Pagliara$^1$, Matthias Hempel$^3$, J\"urgen Schaffner-Bielich$^1$}
\affil{$^1$Institute for Theoretical Physics, Ruprecht-Karls University, Philosophenweg 16, 69120 Heidelberg, Germany}
\affil{$^2$Department of Physics \& Astronomy, Michigan State University, East  Lansing, MI 48824, USA}
\affil{$^3$Department of Physics, University of Basel, Klingelbergstrasse 82, 4056 Basel, Switzerland}
\begin{abstract}
The recent observation of the pulsar PSR J1614-2230 with a mass of 
$1.97\pm0.04\:$M$_\odot$ gives a strong constraint on the quark and 
nuclear matter equations of state (EoS). 
We explore the parameter ranges for a parameterized EoS for quark stars. 
We find that strange stars, made of absolutely stable strange quark matter, 
comply with the new constraint only if effects from the strong coupling 
constant and color-superconductivity are taken into account. 
Hybrid stars, compact stars with a quark matter core and an hadronic 
outer layer, can be as massive as $2\:$M$_\odot$, but only for a 
significantly limited range of parameters. 
We demonstrate that the appearance of quark matter in massive stars depends 
crucially on the stiffness of the nuclear matter EoS. 
We show that the masses of hybrid stars stay below the ones of hadronic and 
pure quark stars, due to the softening of the EoS at the quark-hadron phase transition. 
\end{abstract}
\section{Introduction}
The densities in the interior of neutron stars exceed the ground state density of atomic 
nuclei, $n_0 \sim 0.16\:$fm$^{-3}$, by far. This naturally raises the idea, that compact 
stars might contain a deconfined and chirally restored quark phase. Recently, \cite{Demorest10} 
found a new robust mass limit for compact stars by determining the mass of the millisecond 
pulsar PSR J1614-2230 to be $M=1.97 \pm 0.04\:$M$_\odot$. This value, together with 
the mass of pulsar J1903+0327 of $M=1.667 \pm 0.021\:$M$_\odot$ \citep{Freire10} is 
much larger than the Hulse-Taylor limit of $M\sim 1.44\:$M$_\odot$ \citep{Thorsett99}, 
which for a long time has been the highest precisely measured pulsar mass. In this letter 
we want to explore the implications of this new measurement on the possible presence of 
quark matter in compact stars. Moreover, our aim is to map out the parameter range 
for the widely used quark bag model with respect to its ability to reproduce high mass compact 
stars such as PSR J1614-2230.\\ 
There are two classes of compact stars which contain quark matter. 
The first class are so-called hybrid stars, with quarks only in their interior either in form of a 
pure quark matter core or a quark-hadron mixed phase. The size of the core depends hereby
on the critical density for the quark-hadron phase transition $n_{\it{crit}}$ under neutron star 
conditions. The second class of so-called (strange) quark stars is realized for the special 
scenario of absolutely stable strange quark matter (see e.g. \cite{Itoh70, Bodmer71,Witten84}). 
It is based on the idea that the presence of strange quarks can lower the energy per 
baryon of the mixture of up, down, and strange quarks in weak equilibrium below the 
one of $^{56}$Fe ($\sim 930\:$MeV).  As a consequence, this strange quark matter forms the true 
ground state of nuclear matter and occupies the entire compact star \citep{Alcock86,Haensel86}.\\
The mass measurement for PSR J1614-2230 sets for the first time very strong limits for 
the parameters of any zero temperature equation of state (EoS), and thereby also for the one of quark matter. 
Usually, the appearance of strangeness in quark and hadronic matter provides an additional 
degree of freedom and thereby softens the nuclear EoS, that is, decreases the 
pressure for a given energy density. As a result, quark and hybrid stars cannot reach high masses. 
However, many studies found that effects from the strong interaction, such as 
one-gluon exchange or color-superconductivity can stiffen the quark matter EoS and increase 
the maximum mass of quark and hybrid stars \citep{Ruester04,Horvath04,Alford07, Fischer10, Kurkela10a, Kurkela10b}.
\cite{Ozel10} and \cite{Lattimer10} gave first studies on the implications of the new mass limits from 
PSR J1614-2230 for quark and hybrid stars in the quark bag model. However, as we will show below, a 
systematic analysis of the whole allowed parameter range is still missing.\\
\cite{Lattimer10} include strange quark matter in form of a bag model EoS for quark 
stars as well as hybrid stars. The authors do not study strong effects from color-superconductivity 
and impose the additional constraint of $n_{\it{crit}} \gtrsim n_0$. This is a reasonable \citep{Lattimer10}, 
but not necessary condition \citep{Witten84}. Moreover, they exclude a priori the existence of a 
quark-hadron mixed phase and come to the conclusion, that the existence of a $2.5\:$M$_{\odot}$ star 
would exclude the quark-hadron phase transition in compact star interiors. As we will show in the next 
sections our results cannot confirm this statement, furthermore we find that a quark-hadron mixed phase 
in fact plays a major role in supporting high mass hybrid stars.\\
In a different analysis by \cite{Ozel10}, the authors studied the implications of the new measurement on 
hybrid stars with a parameterized quark bag model including effects from color-superconductivity and QCD 
corrections. They find that both effects are required to support the mass of PSR J1614-2230. However, 
\cite{Ozel10} adjust the bag constant to obtain a fixed density of $n_{\it{crit}}=1.5n_0$ for the phase 
transition to quark matter from the relatively soft APR nucleonic EoS \citep{Akmal98}. 
We find that the stiffness of the hadronic EoS is important for large hybrid star masses and also 
that the maximum mass of hybrid star configurations experiences a minimum at around $n_{\it{crit}}=0.1 - 0.2\:$fm$^{-3}$ 
- values close to the critical density which \cite{Ozel10} choose for their calculations. \\
Therefore, the aim of this paper is to fully and systematically exploit the constraints on 
the quark bag EoS provided by the new mass limit of \cite{Demorest10}. 
Quark matter is described by a bag model EoS with first order corrections from 
the strong interaction coupling constant and effects from finite strange quark mass and
color-superconductivity. For the hybrid star calculations, we use the two different relativistic mean-field (RMF) 
parameter sets TM1 \citep{Sugahara94} and NL3 \citep{Lalazissis97}, to explore the influence of the hadronic 
part of the EoS. In our calculations we do not include hyperons which can alter the quark hadron phase 
transition \citep{Bhattacharyya11}. However, their exact role is currently an open question 
(see e.g. \cite{Yasutake10}). Therefore, in this work, we will focus on non-strange hadronic matter.
We consider the two possible extreme cases for the phase coexistence between quark and hadronic 
matter: the Maxwell transition, corresponding to a very large surface tension of quark matter 
\citep{Heiselberg93}, and the Gibbs construction \citep{Glendenning92} which completely neglects 
Coulomb and surface energies.\\
In the following we will describe our results and compare them with the aforementioned studies. 
Sections \ref{uqm} and \ref{cscqm} are devoted to quark stars with unpaired and 
color-superconducting quark matter in the color-flavor-locked (CFL) phase, respectively. 
Hybrid stars are discussed in section \ref{hybrid}.
\section{Quark stars}
\subsection{Unpaired quark matter}
\label{uqm}
\begin{figure}
\includegraphics[width=8cm]{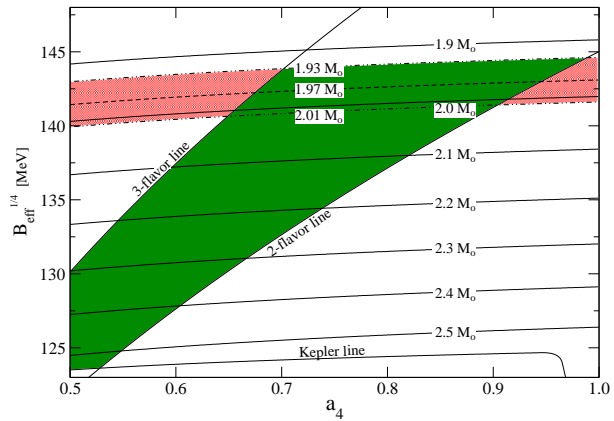}
\caption{Maximum masses of unpaired strange quark stars as a function of $B_{\it{eff}}$ and $a_4$. 
The green shaded area marks the allowed parameter region according to the constraints of the existence 
of nuclei (2-flavor line), absolute stability of strange quark matter (3-flavor line), stability of fast rotating 
stars (Kepler line), and the mass of PSR J1614-2230 including its $1\sigma$ error.}
\label{unpaired}
\end{figure}

For the strange quark matter, we take the modified bag model: 
\begin{equation}
 \Omega_{QM} =  \sum_{i=u,d,s,e} \Omega_i + \frac{3\mu^4}{4 \pi^2}(1-a_4)+ B_{\it{eff}}  ,
\label{eq1}
\end{equation}
where $\Omega_i$ are the Grand potentials for the up, down, and strange quarks and electrons describing these as non-interacting fermions. 
We choose the strange quark mass to be $m_s=100\:$MeV \citep{Amsler10} while the masses of the up and down quarks and electrons are set to zero.
In the sense of the generic quark matter EoS from \cite{Alford05}, we have added the $a_4$ term with the baryon chemical potential $\mu$ of the quarks in order to 
account for corrections from strong interaction. 
The usual approach in quark bag models is to unite all non-perturbative effects of the strong interactions into 
a bag constant $B$. The EoS can then be extended by including first order 
corrections in the strong coupling constant (see e.g. Fraga et al. 2001). The quark bag model in equation (\ref{eq1}) is motivated by this 
approach. However, since quark star matter is not in the perturbative regime, we consider $a_4$ and the bag constant
as effective parameters, denoting the latter by $B_{\it{eff}}$, and explore their whole parameter range. Therefore, we vary $a_4$ from $a_4 = 1$, 
which corresponds to no QCD corrections, to small values when the corrections are strong. Equation (\ref{eq1}) enables us to 
compute the pressure, energy density, and baryon number density assuming charge neutrality and $\beta-$equilibrium. 
By solving the Tolman-Oppenheimer-Volkoff equations we obtain the maximum quark star masses. \\
Following \cite{Farhi84}, we require nonstrange quark matter in bulk to have a binding 
energy per baryon higher than the one of the most stable atomic nucleus, $^{56}$Fe, 
which is $930\:$MeV, plus a $4\:$MeV correction coming from surface effects. By imposing 
that $E/A \geq 934\:$MeV for $2-$flavor quark matter at ground state, we ensure that 
atomic nuclei do not dissolve into their constituent quarks. Thereby we obtain an upper 
limit on the maximum mass of strange quark stars denoted as "2-flavor line" in Fig. \ref{unpaired}.
Another constraint is given by the implementation of the strange matter hypothesis 
\citep{Bodmer71,Witten84} as described in the introduction, with $E/A \leq 930\:$MeV for 
strange quark matter at ground state \citep{Farhi84}. This condition results in the "3-flavor 
line" in Fig. \ref{unpaired} and gives a lower limit on the maximum masses. Fig. 
\ref{unpaired} also shows lines of constant maximum mass. The three dotted lines enclosing 
the red shaded area represent the mass of PSR J1614-2230 with its $1\sigma$ error \citep{Demorest10}. 
The 2-flavor and the 3-flavor lines cross on the left outside the plot range
at $a_4=0.247$, $B_{\it{eff}}^{1/4}=102.24\:$MeV 
which correspond to a maximum mass star with M$=3.36\:$M$_\odot$ and a radius of $19\:$km. 
The Kepler line at low $B_{\it{eff}}$ represents a limit
for quark stars which can rotate with a Keplerian frequency of at least 716 Hz \citep{Hessels06}. 
Therefore, the green shaded area is the allowed quark star parameter region with a maximum 
mass of $2.54\:$M$_\odot$ at $a_4\approx0.53$ and 
$B_{\it{eff}}^{1/4}\approx123.7\:$MeV. 
However, the Kepler line is obtained from a parametrization of 
\cite{Lattimer:2006xb} and gives a rough estimate when applied to strange stars. 
For a more reliable Kepler limit, 
the presented quark EoSs should be applied in general relativistic calculations 
of rotating quark stars similar to the studies of \cite{Haensel09} or \cite{Lo11}. 
From Fig.\ref{unpaired}, we see that for $a_4=1$ the 2-flavor line requires 
M$_{\it{max}}\lesssim 1.92\:$M$_\odot$ which is ruled out by the new mass limit, at 
least within its $1\sigma$ error. Thus we find that $a_4<1$, i.e. QCD corrections 
must be included to ensure the compatibility of the model with observational data. 
\subsection{Color-superconducting strange matter stars}
\label{cscqm}
\begin{figure}
\includegraphics[width=8cm]{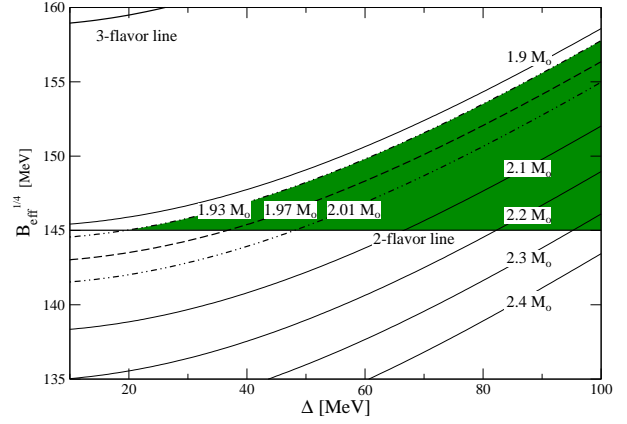}
\caption{Same as in Fig. \ref{unpaired}, but this time for color-superconducting strange quark 
matter with the pairing gap $\Delta$ and fixed $a_4=1.0$.}
\label{CFL}
\end{figure}

At large densities, such as in compact star interiors, up, down, and strange quarks are assumed 
to undergo pairing and form the so-called CFL phase. We adopt the 
EoS from \cite{Alford01} which introduces the pairing energy $\Delta$ as a new free parameter:
\begin{eqnarray}
\Omega_{\it{CFL}}&=&\frac{6}{\pi^2}\int_0^{\nu}{\it{dp}}\:p^2(p-\mu)+\frac{3}{\pi^2}\int_0^{\nu}{\it{dp}}\:p^2(\sqrt{p^2+m_s^2}-\mu)\nonumber\\
&&+(1-a_4)\frac{3\mu^4}{4\pi^2}-\frac{3\Delta^2\mu^2}{\pi^2}+B_{\it{eff}}\label{eq2}   ,
\end{eqnarray}
where $\nu=2\mu-\sqrt{\mu^2-m_s^2/3}$. 
We added again the $a_4-$term to account for QCD corrections. 
The results for $a_4=1.0$ and $m_s=100\:$MeV are shown in 
Fig. \ref{CFL} where we have imposed the same constraints as in Fig. \ref{unpaired}. 
The "3-flavor" and "2-flavor" lines give again a lower and upper limit on 
the maximum mass. The green shaded area is the maximally allowed parameter region. 
Its upper edge is solely given by the constraint from the mass measurement of PSR J1614-2230. Note, that for 
small values of $\Delta$ quark stars are not allowed. For large gaps starting at 
$\Delta \gtrsim 20\:$MeV the allowed area opens up and one can obtain high maximum masses. 
The largest mass allowed within the plot range is 
$2.34\:$M$_\odot$ at $a_4=1.0$, $B_{\it{eff}}^{1/4}=145\:$MeV and $\Delta=100\:$MeV. 
If we assume the same constraint from the Keplerian frequency as before, we find, that it 
has no influence on our results within the plot range of Fig. \ref{CFL}. More exotic 
parameter combinations, as e.g. $a_4=0.66$, $B_{\it{eff}}^{1/4}=130.5\:$MeV, $\Delta=50\:$MeV 
and $a_4=0.75$, $B_{\it{eff}}^{1/4}=134.9\:$MeV, $\Delta=100\:$MeV with maximum masses of 
$2.5\:$M$_\odot$ and $2.8\:$M$_\odot$ respectively are also allowed.\\
Taking into account QCD corrections, i.e lowering $a_4$, gives maximum mass lines shifted 
to only slightly lower values of $B_{\it{eff}}$. Together with the lowered 2-flavor and 
3-flavor lines this means that at some point the 3-flavor constraint will become important. 
Still we can obtain quite large maximum masses at a sufficiently high gap value. 
Varying the strange quark mass basically results in shifting the maximum mass lines along 
the gap-axis as the crucial contribution to the EoS comes from 
a term $m_s^2-4\Delta^2$ \citep{Alford05}. 
Thus, for fixed $a_4$ and a higher strange mass, the allowed area opens up at a larger value 
of $\Delta$.
\section{Hybrid stars}
\label{hybrid}
\begin{figure}
\includegraphics[width=6cm, angle=270]{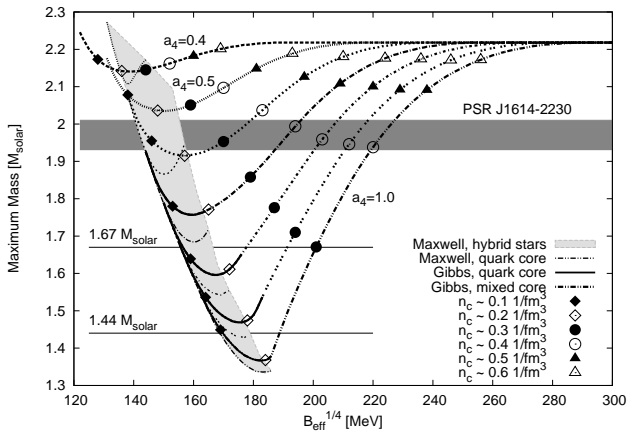}
\caption{Maximum masses and critical densities of hybrid stars calculated with the 
TM1 RMF EoS as a function of $B_{\it{eff}}$ and $a_4$. The quark-hadron phase transitions are modelled by the Maxwell and Gibbs approach. 
For the latter, solid lines indicate pure quark matter cores and dashed lines indicate mixed phase cores.}
\label{mass_tm1}
\end{figure}

\begin{figure}
\includegraphics[width=6cm, angle=270]{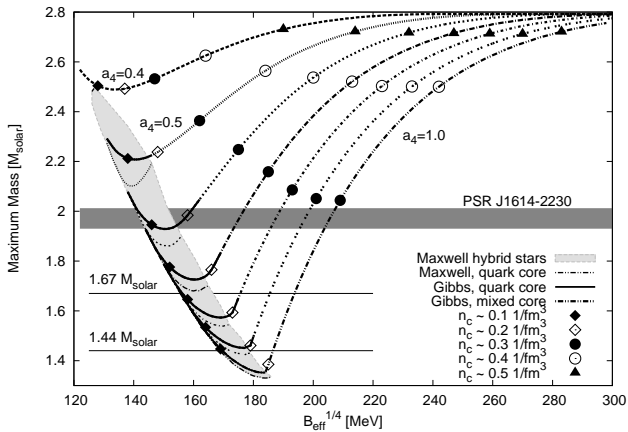}
\caption{Same as in Fig. \ref{mass_tm1} with hadronic matter described by the NL3 RMF EoS.}
\label{mass_nl3}
\end{figure}

For the hybrid star calculations we use again the bag model EoS of equation (\ref{eq1}) 
and the same quark masses as in the previous sections. As hadronic EoSs 
we choose the TM1 and NL3 RMF parameter sets, with maximum 
neutron star masses of $2.2\:$M$_\odot$ and $2.78\:$M$_\odot$, respectively 
\citep{Sugahara94,Lalazissis97}. The critical density $n_{\it{crit}}$ for the quark-hadron 
phase transition is mainly given by the effective bag constant $B_{\it{eff}}$ while corrections 
from the strong interaction $a_4$ affect the stiffness of the quark EoS. Combinations of 
small values for $a_4$ and $B_{\it{eff}}$ lead therefore to stiff hybrid EoSs with a low 
critical density \citep{Fischer10}. As a consequence, the corresponding hybrid stars can 
be very massive and have a large quark matter core. Similar to \cite{Ozel10} we omit all
parameter combinations which lead to several transitions between quark and hadronic matter, 
which happens for small $B_{\it{eff}}$ and $a_4$.\\
The phase transitions are modelled by the Gibbs and Maxwell constructions (see e.g. \cite{Glendenning92}) 
corresponding to a small (Gibbs) and to a large value of the surface tension (Maxwell) (see discussion in \cite{Page:2006ud}). 
The former leads to the presence of an extended quark-hadron mixed phase region in the compact star 
interior where the pressure rises smoothly with density. With the Maxwell approach, matter experiences 
a direct transition from hadronic to quark matter in cold compact stars, accompanied by a density jump from lower 
(hadronic phase) to higher (quark phase) densities. Applying both models for the phase transition 
we calculate the hybrid star maximum masses and plot them
as a function of the bag constant $B_{\it{eff}}$ for fixed values of $a_4$. Fig. \ref{mass_tm1} 
shows the hybrid star maximum mass curves for the TM1 EoS for the 
hadronic phase, while in Fig. \ref{mass_nl3} we show the results for the NL3 model. The 
lines which extend from low values to high values of $B_{\it{eff}}$ correspond to calculations 
using the Gibbs phase transition. Stars on the solid lines have a pure quark matter core while the 
dashed lines represent stars where only a mixed phase is present.
The maximum masses for the Maxwell transition are represented by the grey shaded area. 
Due to the absence of a mixed phase, hybrid stars in the Maxwell approach can only 
contain a pure quark matter core. A too large density jump from hadronic to quark matter 
in their interior leads to a gravitational instability against radial oscillations. 
As a consequence we find from Figs. \ref{mass_tm1} and \ref{mass_nl3} that the 
parameter range for hybrid stars in the Maxwell approach is significantly reduced in 
comparison to the "Gibbs hybrid stars". Furthermore, in most of the cases we find that for the 
same combinations of $B_{\it{eff}}$ and $a_4$ the Maxwell phase transition leads to 
lower hybrid star maximum masses than the ones with a Gibbs construction.
Only for low values of $a_4$ when quark matter becomes very stiff, the maximum masses of the 
"Maxwell hybrid stars" can become significantly larger than their Gibbs counterparts. 
\newline
For stable hybrid stars with the Maxwell and Gibbs transition, the pure quark and the mixed phase 
extend over almost the whole star when $B_{\it{eff}}$, i.e. the critical density $n_{\it{crit}}$, is low.
However, as can be seen from Figs. \ref{mass_tm1} and \ref{mass_nl3}, a pure quark matter core in 
hybrid stars occurs only for small values of $B_{\it{eff}}$. For the TM1 hadronic EoS, 
sufficiently massive Gibbs hybrid stars contain only a mixed phase in the core. On the other hand, for the NL3 hadronic EoS with 
low values of $B_{\it{eff}}$ and $n_{\it{crit}}$, large cores of pure quark matter exist down to $a_4 \sim 0.5$. 
This is a consequence of the much stiffer NL3 hadronic EoS in comparison to the TM1 parameter set. 
Nevertheless, hybrid stars with a pure quark core and a mass $\ge 1.93\:$M$_\odot$ are only obtained for 
$a_4\lesssim 0.6$.\\
A common feature which is seen for hybrid stars in the Gibbs approach is that for a 
fixed value of $a_4$ the maximum masses decrease with lower critical densities and experience
a minimum around $n_{\it{crit}} \sim 0.2\:$fm$^{-3}$. For smaller 
$n_{\it{crit}}$, the quark EoS starts to dominate and the maximum masses increase again as they 
approach the limit of absolutely stable strange quark matter.
We plot the mass of PSR J1614-2230 with its $1\sigma$ error 
as a gray band as well as lines at M$=1.44\:$M$\odot$ and M$=1.67\:$M$\odot$ to indicate 
the masses of the Hulse-Taylor pulsar and of J1903+0327 respectively \citep{Thorsett99,Freire09}.
Especially in Fig. \ref{mass_tm1} it can be seen that the new mass limit significantly 
tightens the constraints on the model parameters, as the whole area below the gray band is 
now excluded while formerly this was only the case below the $1.67\:$M$_\odot$ line.
In this work we did not consider effects of a finite surface tension on the quark-hadron 
mixed phase \citep{Heiselberg93} which has been found to be an intermediate of the Gibbs 
and the Maxwell constructions \citep{Maruyama07, Endo06}. However, the $B_{\it{eff}}-a_4$ 
parameter space for stable hybrid stars with finite surface tension can be expected to be between the Gibbs 
and the Maxwell constructions. Therefore our result, that hybrid stars can be massive, remains valid.
\section{Conclusion}
We present for the first time a comprehensive and systematic study on the constraints of the 
new compact star mass limit from the millisecond pulsar PSR J1614-2230 
on the properties of quark and hybrid stars modelled within an extended quark bag model. 
The parameters of the bag model are an effective bag constant, corrections from the strong 
interaction coupling constant and color-superconductivity. 
We find that the new mass limit does not rule out the possibility of having quark matter in compact 
stars but provides tight bounds on its properties. 
High compact star masses where 
quark matter is the dominant component, require strong QCD corrections and/or a large 
contribution from color-superconductivity. 
In this case strange stars can reach masses far beyond $2\:$M$_\odot$. 
For hybrid stars with a sizeable quark matter phase, our investigation shows that 
pure quark matter cores are obtained only for a small parameter range when the hadronic 
EoS is stiff and the critical density for the quark-hadron phase transition is around saturation density. 
Our results agree with \cite{Ozel10} concerning the importance of effective 
QCD corrections to reach high compact star masses. Contrarily to one of the statements of \cite{Lattimer10}, 
we find that pairing helps to increase the maximum mass and that corrections from the strong interaction have a significant effect. 
For hybrid stars we demonstrate that the allowed parameter region hinges crucially on the 
stiffness of the hadronic EoS and can therefore be much larger than in the case of \cite{Ozel10}.
Recently, \cite{vanKerkwijk:2010mt} reported the possible existence of a $2.4\:$M$_\odot$ star. If 
such a measurement is confirmed, even larger corrections from the strong coupling constant 
and larger values of the CFL gap will be required for strange stars. Hybrid stars 
could exist only for a stiff hadronic EoS and would contain only a core with 
a quark-hadron mixed phase in our approach.\\
An investigation similar to the one we presented here would also be desirable for other effective models of QCD such as 
the Nambu-Jona-Lasinio (NJL) and the PNJL models \citep{Klahn:2006iw,Pagliara:2007ph,Blaschke10,Ippolito:2007hn}, as well as
Schwinger-Dyson approaches, as recently proposed by \cite{Li:2011vd}. Although mass measurements are very useful for 
constraining the nuclear matter EoS, additional observational information is required to probe the existence of quark 
matter in compact stars. Cooling, r-modes calculations, and gravitational wave signals of mergers \citep{Bauswein09} 
are promising tools. Also heavy ions collisions experiments provide crucial
information on the nuclear matter EoS: an extended analysis on the
quark models paramaters which includes both astrophysical constraints,
as the mass of PSR J1614-2230, and terrestrial laboratories constraints
would be extemely interesting.\\
\textit{Acknowledgments} 
I.S. is supported by the Alexander von Humboldt foundation via a Feodor Lynen fellowship 
and wishes to acknowledge the support of the Michigan State University High Performance Computing Center 
and the Institute for Cyber Enabled Research. 
G.P. acknowledges support from the Deutsche Forschungsgemeinschaft (DFG) under grant no. 
PA 1780/2-1 and the work of J.S.-B. is supported by the DFG through the Heidelberg Graduate 
School of Fundamental Physics. M.H. acknowledges support from the High Performance and High 
Productivity Computing (HP2C) project. This work is supported by BMBF under grant FKZ 06HD9127, 
by the Helmholtz Alliance HA216/EMMI and by CompStar, a research networking program of the 
European Science Foundation.\\

\end{document}